\documentclass[aps,floatfix,superscriptaddress,preprint]
{revtex4}

\usepackage{titlesec}
\usepackage{amssymb,amsmath,amsfonts,latexsym,graphicx,epsfig,bm}
\usepackage{epstopdf}

\titleformat{\section}{\large\bfseries}{\thesection}{1em}{}

\newcommand{\bea}{\begin{eqnarray}}
\newcommand{\ena}{\end{eqnarray}}
\newcommand{\nn}{\nonumber\\}
\newcommand{\be}{\begin{equation}}
\newcommand{\en}{\end{equation}}

\newcommand{\ed}{\end{document}}

\newcommand{\slp}{p\kern-5pt/}
\newcommand{\Tr}{\mbox{\rm{tr}}}

\begin{document}

\title{Exclusive decays \boldmath{$J/\psi \to D_{(s)}^{(*)-} {\ell}^+ \nu_{\ell}$}\\
in a covariant constituent quark model
 with infrared confinement}

\author {M. A. Ivanov}
\email{ivanovm@theor.jinr.ru}
\affiliation{
Bogoliubov Laboratory of Theoretical Physics,
Joint Institute for Nuclear Research,
141980 Dubna, Russia}

\author{C. T. Tran}
\email{ctt@theor.jinr.ru,tranchienthang1347@gmail.com}
\affiliation{
Bogoliubov Laboratory of Theoretical Physics,
Joint Institute for Nuclear Research,
141980 Dubna, Russia}
\affiliation{Advanced Center for Physics, Institute of Physics, Vietnam 
Academy of Science and Technology, 100000 Hanoi, Vietnam}
\affiliation{Department of General and Applied Physics, 
Moscow Institute of Physics and Technology, 141700 Dolgoprudny, Russia}
\date{\today}

\begin{abstract}
We investigate the exclusive semileptonic decays 
$J/\psi \to D_{(s)}^{(*)-} {\ell}^+ \nu_{\ell}$, where $\ell=e,\mu$, within 
the Standard Model. The relevant transition form factors are calculated in 
the framework of a relativistic constituent quark model with built-in 
infrared confinement. Our calculations predict the branching fractions 
$\mathcal{B}(J/\psi \to D_{(s)}^{(*)-} {\ell}^+ \nu_{\ell})$ to be of the order 
of $10^{-10}$ for $D_s^{(*)-}$ and $10^{-11}$ for $D^{(*)-}$. Most of our numerical 
results are consistent with other theoretical studies. However, some branching 
fractions are larger than those calculated in QCD sum rules approaches 
but smaller than those obtained in the covariant light-front quark model 
by a factor of about $2-3$.
\end{abstract}

\maketitle


\section{\label{sec:intro}introduction}

Low lying states of quarkonia systems similar to $J/\psi$ usually decay 
through intermediate photons or gluons produced by the parent $q\bar{q}$ quark 
pair annihilation~\cite{Sharma:1998gc}. As a result, strong and 
electromagnetic decays of $J/\psi$ have been largely investigated while weak 
decays of $J/\psi$ have been put aside for decades. However, in the last 
few years many improvements in instruments and experimental techniques, 
in particular, the luminosity of colliders,  have led to observation of many  
rare processes including the extremely rare decays $B_{(s)}^0 \to \mu^+\mu^-$, 
announced lately by the CMS and LHCb collaborations~\cite{CMS:2014xfa}. 
The branching fractions were measured to be 
$\mathcal{B}(B_s^0 \to \mu^+\mu^-) = (2.8^{+0.7}_{-0.6}) \times 10^{-9}$ and 
$\mathcal{B}(B^0 \to \mu^+\mu^-) = (3.9^{+1.6}_{-1.4}) \times 10^{-10}$. 
This raises the hope that one may also explore the rare weak decays of 
charmonium and draws researchers' attention back to these modes.

Recently, BESIII Collaboration reported on their search for semileptonic weak 
decays $J/\psi \to D^{(*)-}_s e^+ \nu_e + \textrm{c.c.}$~\cite{Ablikim:2014fpb}, 
where ``$+\textrm{c.c.}$'' indicates that the signals were sum of these modes and 
the relevant charge conjugated ones. The results at $90\%$ confidence level 
were found to be 
$\mathcal{B}(J/\psi \to D^-_s e^+ \nu_e + \textrm{c.c.}) < 1.3\times10^{-6}$ and 
$\mathcal{B}(J/\psi \to D^{*-}_s e^+ \nu_e + \textrm{c.c.}) < 1.8\times10^{-6}$. 
Although these upper limits are far above the predicted values within the 
Standard Model (SM), which are of the order of 
$10^{-8}-10^{-10}$~\cite{SanchisLozano:1993ki, Wang:2007ys, Shen:2008zzb}, 
one should note that this was the first time an experimental constraint on 
the branching fraction $\mathcal{B}(J/\psi \to D^{*-}_s e^+ \nu_e + \textrm{c.c.})$ 
was set, and moreover, the constraint on the branching fraction 
$\mathcal{B}(J/\psi \to D^-_s e^+ \nu_e + \textrm{c.c.})$ was $30$ times more 
stringent than the previous one~\cite{Agashe:2014kda}. With a huge data 
sample of $10^{10}$ $J/\psi$ events accumulated each year, BESIII is expected 
to detect these decays, even at SM levels, in the near future. 

From the theoretical point of view, these weak decays are of great importance 
since they may lead to better understanding of nonperturbative QCD effects 
taking place in transitions of heavy quarkonia. Moreover, the semileptonic 
modes $J/\psi \to D^{(*)}_{(s)} \ell\nu$, as three-body weak decays of a vector 
meson, supply plentiful information about the polarization observables that 
can be used to probe the hidden structure and dynamics of hadrons. 
Additionally, these decays may also provide some hints of new physics beyond 
the SM, such as TopColor models~\cite{Hill:1994hp}, the Minimal Supersymmetric 
Standart Model (MSSM) with or without R-parity~\cite{Martin:1997ns}, and 
the two-Higgs-doublet models (2HDMs)~\cite{Hou:1992sy, Baek:1999ch}.

The very first estimate of $\mathcal{B}(J/\psi \to D^{(*)}_s \ell \nu)$ was 
made based on the (approximate) spin symmetry of heavy mesons, giving an 
inclusive branching fraction of $(0.4 - 1.0)\times 10^{-8}$, summed over 
$D_s$, $D_s^*$, $e$, $\mu$ and both charge conjugate 
modes~\cite{SanchisLozano:1993ki}. In this work the transition form factors 
were parametrized through a universal function, similar to the Isgur-Wise 
function in the heavy quark limit. However, the zero-recoil approximation adopted 
in calculating the hadronic matrix elements led to large uncertainties in the 
decay width evaluation. For that reason, author of~\cite{SanchisLozano:1993ki}
noted that these results should be viewed as an estimate suggesting 
experimental searching, rather than a definite prediction. Recently, 
by employing QCD sum rules (QCD SR)~\cite{Wang:2007ys} or making use of 
the covariant light-front quark model (LFQM)~\cite{Shen:2008zzb}, new 
theoretical studies found the branching fractions of 
$J/\psi \to D^{(*)-}_s e^+ \nu_e \,+\, c.c.$ to be of the order of $10^{-10}$. 
However, the results presented in~\cite{Shen:2008zzb} were about $2-8$ times 
larger than those calculated in~\cite{Wang:2007ys}. Besides, one can 
significantly reduce hadronic uncertainties and other physical constants like 
$G_F$ and $|V_{cs}|$ by considering the ratio of branching fractions 
$R \equiv 
\mathcal{B}(J/\psi \to D^*_s \ell \nu)/\mathcal{B}(J/\psi \to D_s \ell \nu)$. 
This ratio had been predicted to be $\simeq1.5$ in~\cite{SanchisLozano:1993ki} 
while the recent study~\cite{Wang:2007ys} suggested $R \simeq 3.1$. Clearly, 
more theoretical studies and cross-check are necessary.

In the present work we offer an alternative approach to the investigation of 
the exclusive decays $J/\psi \to D_{(s)}^{(*)-} {\ell}^+ \nu_{\ell}$, in which we 
employ the covariant constituent quark model with built-in infrared confinement
 [for short, confined covariant quark model (CCQM)] as dynamical input to 
calculate the nonperturbative transition matrix elements. Our paper is organized as follows: In Sec.~\ref{sec:model}, we set up our framework by briefly introducing the CCQM. Sec.~\ref{sec:matrixElement} contains the definitions and derivations of the form factors of the decays $J/\psi \to D_{(s)}^{(*)-} {\ell}^+ \nu_{\ell}$ based on the effective Hamiltonian formalism. In this section we also describe in some detail how calculation of the form factors proceeds in our approach. Sec.~\ref{sec:FF} is devoted to the numerical results for the form factors, including comparison with the available data. Sec.~\ref{sec:branching} contains our numerical results for the branching fractions. And finally, we make a brief summary of our main results in Sec.~\ref{sec: summary}.

\section{model}
\label{sec:model}

The CCQM has been developed in some of our earlier papers 
(see~\cite{Ivanov:2011aa} and references therein). In the CCQM framework one 
starts with an effective Lagrangian describing the coupling of a meson $H$ to 
its constituent quarks $q_1$ and $q_2$,
\be
\mathcal{L}_{int}(x) = g_H H(x) \int dx_1 \int dx_2 
F_H(x;x_1,x_2)[\bar{q}_2(x_2) \Gamma_H q_1(x_1)]+\textrm{H.c.},
\label{eq:Model_Lagrangian}
\en
where $\Gamma_H$ is the relevant Dirac matrix and $g_H$ is the coupling 
constant. The vertex function $F_H$ is related to the scalar part of the 
Bethe-Salpeter amplitude and characterizes the finite size of the meson. 
Transitions between mesons are evaluated by one-loop Feynman diagrams with 
free quark propagators.The high energy divergence of quark loops is tempered 
by nonlocal Gaussian-type vertex functions with a falloff behavior. 
We adopt the following form,
\be
F_H (x;x_1,x_2)=\delta(x - w_1 x_1 - w_2 x_2) \Phi_H((x_1-x_2)^2),
\en 
where $w_i = m_{q_i}/(m_{q_1}+m_{q_2})$. This form of $F_H$ is invariant under the 
translation $F_H(x+a;x_1+a,x_2+a)=F_H(x;x_1,x_2)$, which is necessary for the 
Lorence invariance of the Lagrangian~(\ref{eq:Model_Lagrangian}).

We adopt a Gaussian form for the vertex function: 
\be
\widetilde\Phi_H(-p^2) = \int\! dx\, e^{ipx} \Phi_H(x^2)
= e^{p^2/\Lambda^2_H}.
\label{eq:Gauss}
\en
The parameter $\Lambda_H$ characterizes the size of the  meson.
The calculations of the Feynman diagrams proceed
in the Euclidean region where $p^2=-p^2_E$ and therefore
the vertex function has the appropriate falloff
behavior to provide for the ultraviolet convergence of the loop integral.
 
\begin{figure}[htbp]
\includegraphics[width=0.40\textwidth]{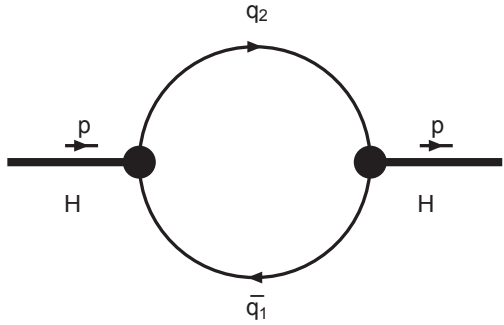}
\caption{One-loop self-energy diagram for a meson.}
\label{fig:mass}
\end{figure}

The normalization of particle-quark vertices is provided by the compositeness 
condition~\cite{Z=0}
\be
Z_H = 1 - \Pi^\prime_H(m^2_H) = 0,
\label{eq:compositeness}
\en
where $Z_H$ is the wave function renormalization constant of the meson $H$ 
and $\Pi^\prime_H$ is the derivative of the meson mass function. To better 
understand the physical meaning of the compositeness condition we want to 
remind the reader that the constant $Z^{1/2}_H$ can be view as the matrix 
element between the physical particle state and the corresponding bare state. 
The compositeness condition $Z_H = 0$ implies that the physical bound state 
does not contain the bare state. The constituents are virtual and they are 
introduced to realize the interaction described by 
the Lagrangian~(\ref{eq:Model_Lagrangian}). As a result of the interaction, 
the physical particle becomes dressed and its mass and wave function are 
renormalized. Technically, the compositeness condition allows one to evaluate 
the coupling constant $g_H$. The meson mass function 
in~(\ref{eq:compositeness}) is defined by the Feynman diagram shown
in Fig.~\ref{fig:mass}. It has the explicit form 
\be
\Pi_P(p) = 3g_P^2 \int\!\! \frac{dk}{(2\pi)^4i}\widetilde\Phi^2_P \left(-k^2\right)
\Tr\left[ S_1(k+w_1p)\gamma^5 S_2(k-w_2p)\gamma^5 \right],
\label{eq:Pmass-Pseudoscalar}
\en  
and
\be
\Pi_V(p) = g_V^2 \left[g^{\mu\nu} - \frac{p^{\mu}p^{\nu}}{p^2}\right] 
\int\!\! \frac{dk}{(2\pi)^4i}\widetilde\Phi^2_V \left(-k^2\right)
\Tr\left[ S_1(k+w_1p)\gamma_{\mu} S_2(k-w_2p)\gamma_{\nu} \right],
\en
for a pseudoscalar meson and a vector meson, respectively. 
Note that we use the free quark propagator
\be
S_i(k) = \frac{1}{m_{q_i} - \not\! k - i\epsilon},
\label{eq:prop}
\en
where $m_{q_i}$ is the constituent quark mass. 

The confinement of quarks is embedded in an effective way: first, 
by introducing a scale intergration in the space of $\alpha$-parameters; 
and second, by truncating this scale intergration on the upper limit that 
corresponds to an infrared cutoff. By doing this one removes all possible 
thresholds in the quark diagram. The cutoff parameter is taken to be universal.
Other model parameters are adjusted by fitting to available experimental data. 
Once these parameters are fixed, one can employ the CCQM as a frame-independent
tool for hadronic calculation. One of the advantages of the CCQM is that in 
this framework the full physical range of momentum transfer is available, 
making calculation of hadronic quantities straightforward without any 
extrapolation.

\section{hadronic matrix elements}
\label{sec:matrixElement}

The effective Hamiltonian describing the semileptonic decays 
$J/ \psi \to D^{(*)-}_{(s)} \ell^+ \nu_{\ell}$ is given by
\be
\mathcal{H}_{\textrm{eff}} (c \to q \ell^+ \nu_{\ell}) = 
\frac{G_F}{\sqrt{2}} V_{cq} 
\left[ \bar{q} O_\mu  c \right] 
\left[ \bar{\nu}_{\ell} O^\mu  \ell \right],
\label{eq:Hamiltonian}
\en
where $q = s, d$, and $O^\mu=\gamma^\mu(1-\gamma_5)$ is 
the weak Dirac matrix with left chirality.

\begin{figure}[htbp]
\includegraphics[width=0.40\textwidth]{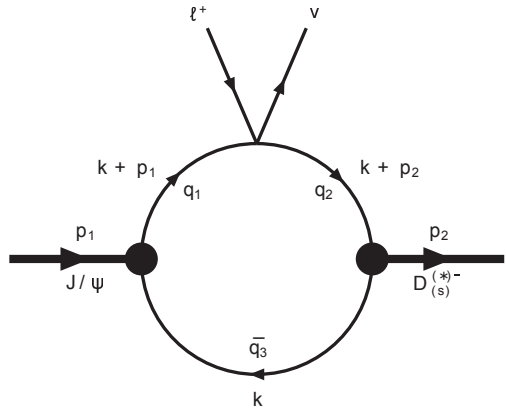}
\caption{Diagram for $J/\psi$ meson semileptonic decays.}
\label{fig:semilept}
\end{figure}

In the CCQM the hadronic matrix elements of the semileptonic $J/\psi$ meson 
decays are defined by the diagram in Fig.~\ref{fig:semilept} and are given by
\bea
&&
\left\langle D^-_{(s)} (p_2) 
\left|\bar{q} O_{\mu}  c \right| J/\psi (\epsilon_1 , p_1) \right\rangle 
=\epsilon_1^{\alpha} T^{\rm VP}_{\mu\alpha}
\nn
 T^{\rm VP}_{\mu\alpha} &=& 3 g_{J/\psi} g_P \int\!\! \frac{d^4k}{(2\pi)^4 i}
\widetilde\Phi_{J/\psi}[-(k + w_{13}p_1)^2] 
\widetilde\Phi_P[-(k + w_{23}p_2)^2]
\nn
&\times& \Tr\left[S_2(k+p_2) O_{\mu} S_1(k+p_1)\gamma_\alpha S_3(k)
\gamma_5\right],
\label{eq:VP}\\[1.2ex]
&&
\left\langle D^{*-}_{(s)} (\epsilon_2, p_2) 
\left|\bar{q} O_\mu  c \right| J/\psi (\epsilon_1 , p_1) \right\rangle
=\epsilon_1^{\alpha}\epsilon_2^{\ast\beta} T^{\rm VV}_{\mu\alpha\beta}
\nn
 T^{\rm VV}_{\mu\alpha\beta} &=&
3 g_{J/\psi} g_V \int\!\! \frac{d^4k}{(2\pi)^4 i}
\widetilde\Phi_{J/\psi}[-(k + w_{13}p_1)^2] 
\widetilde\Phi_V[-(k+w_{23}\,p_2)^2]
\nn
&\times& \Tr\left[ S_2(k+p_2) O_{\mu} S_1(k+p_1)\gamma_\alpha 
S_3(k)\gamma_\beta \right].
\label{eq:VV}
\ena
We use the on-shell conditions $\epsilon_1\cdot p_1 = 0$, 
$\epsilon^\ast_2\cdot p_2 = 0$, and $p_i^2=m_i^2$. Because there are three 
quark types involved in the transition, we have introduced a two-subscript 
notation $w_{ij}=m_{q_j}/(m_{q_i}+m_{q_j})$ $(i,j=1,2,3)$ such that 
$w_{ij}+w_{ji}=1$. 

The loop integrations in Eqs.~(\ref{eq:VP}) and~(\ref{eq:VV}) are done with the help of the Fock-Schwinger 
representation of the quark propagator
\bea
S_q (k + p) &=& \frac{1}{ m_q-\not\! k- \not\! p } 
=  \frac{m_q + \not\! k +  \not\! p}{m^2_q - (k+ p)^2}
\nn
&=& (m_q + \not\! k +  \not\! p)\int\limits_0^\infty \!\!d\alpha 
e^{-\alpha [m_q^2-(k+p)^2]},
\label{eq:Fock}
\ena
where $k$ is the loop momentum and $p$ is the external momentum.
As described later on, the use of the Fock-Schwinger representation allows 
one to do tensor loop integrals in a very efficient way since one can
convert loop momenta into derivatives of the exponent function. 

All loop integrations are performed in Euclidean space. 
The transition from Minkowski space to Euclidean space is performed
by using the Wick rotation
\be
k_0=e^{i\frac{\pi}{2}}k_4=ik_4
\label{eq:Wick}
\en
so that $k^2=k_0^2-\vec{k}^2=-k_4^2-\vec{k}^2=-k_E^2 \leq 0.$
Simultaneously one has to rotate all external momenta, i.e.
 $p_0 \to ip_4$ so that $p^2=-p_E^2 \leq 0$.
Then the quadratic form in Eq.~(\ref{eq:Fock}) becomes positive definite,
\[
m^2_q-(k+p)^2=m^2_q + (k_E+p_E)^2>0,
\]
and the integral over $\alpha$ is absolutely convergent.
We will keep the Minkowski notation to avoid
excessive relabeling. We simply imply that
 $k^2 \leq 0$ and $p^2 \leq 0$.

Collecting the representations for the vertex functions
and quark propagators given by Eqs.~(\ref{eq:Gauss})
and (\ref{eq:Fock}), respectively, one can perform the Gaussian
integration in the expressions for  the matrix elements
in Eqs.~(\ref{eq:VP}) and ~(\ref{eq:VV}). 
The exponent has the form $ak^2+2kr+z_0$, where $r=bp$. 
Using the following properties, 
\be
\left.
\begin{aligned}
    k^\mu\, \exp(ak^2+2kr+z_0) &=\frac{1}{2}\frac{\partial }
{\partial r_\mu }\exp(ak^2+2kr+z_0)\\
    k^\mu k^\nu\, \exp(ak^2+2kr+z_0) &=
      \frac{1}{2}\frac{\partial }{\partial r_\mu } 
      \frac{1}{2} \frac{\partial }{\partial r_\nu }         \exp(ak^2+2kr+z_0)
\\
\text{etc.}&
\end{aligned}
\right\},
\label{eq:change-to-r}
\en
one can replace
$\not\! k $ by $ {\not\! \partial}_r 
= \gamma^\mu\frac{\partial}{\partial r_\mu}$
which allows one to exchange the tensor integrations
for a differentiation of the Gaussian exponent $e^{-r^2/a}$
which appears after integration over loop momentum. 
The $r$-dependent Gaussian exponent $e^{-r^2/a}$ can be moved to the left 
through the differential operator $\not\! \partial_r$ by using the following 
properties,
\bea
\frac{\partial}{\partial r_\mu}e^{-r^2/a} &=& e^{-r^2/a}
\left[-\frac{2r^\mu}{a}+\frac{\partial}{\partial r_\mu}\right],
\nn[1.2ex]
\frac{\partial}{\partial r_\mu}
\frac{\partial}{\partial r_\nu}e^{-r^2/a} &=& e^{-r^2/a}
\left[-\frac{2r^\mu}{a}+\frac{\partial}{\partial r_\mu}\right]\cdot
\left[-\frac{2r^\nu}{a}+\frac{\partial}{\partial r_\nu}\right],
\nn[1.2ex]
\text{etc.}&&
\label{eq:dif}
\ena
Finally, one has to move the derivatives to the right by using
the commutation relation
\be
\left[\frac{\partial}{\partial r_\mu},r^\nu \right]
=g^{\mu\nu}.
\label{eq:comrel}
\en
The last step has been done by using a \textsc{form} code which
works for any numbers of loops and propagators.
In the remaining integrals over the Fock-Schwinger parameters 
$0\le \alpha_i<\infty$
we introduce an additional integration which converts the set of 
Fock-Schwinger parameters into a simplex. We use the transformation
\be
\prod\limits_{i=1}^n\int\limits_0^{\infty} 
\!\! d\alpha_i f(\alpha_1,\ldots,\alpha_n)
=\int\limits_0^{\infty} \!\! dtt^{n-1}
\prod\limits_{i=1}^n \int\!\!d\alpha_i 
\delta\left(1-\sum\limits_{i=1}^n\alpha_i\right)
  f(t\alpha_1,\ldots,t\alpha_n).
\label{eq:simplex}  
\en

The integral over $t$ is well defined and convergent 
below the threshold  $p_1^2< (m_{q_c} + m_{q})^2$.  
The convergence of the integral above  threshold  
$p_1^2\ge (m_c + m_{q})^2$ is guaranteed by the addition of a small imaginary 
to the quark mass, i.e. $m_q\to m_q - i\epsilon, \epsilon>0$
in the quark propagator. It allows one to rotate
the integration variable $t$ to the imaginary axis $t\to i t$. 
As a result the integral  becomes convergent but obtains an imaginary part 
corresponding to quark pair production.

However, by cutting the scale integration at the upper limit 
corresponding to the introduction of an infrared cutoff
\be
\int\limits_0^\infty dt (\ldots) \to \int\limits_0^{1/\lambda^2} dt (\ldots).
\label{eq:conf}
\en
one can remove all possible thresholds present in the initial quark
diagram~\cite{Branz:2009cd}. Thus the infrared cutoff parameter 
$\lambda$ effectively guarantees the confinement of quarks within hadrons. 
This method is quite general and can be used for diagrams with an arbitrary 
number of loops and propagators. 
In the CCQM the infrared cutoff parameter $\lambda$ is taken to be universal 
for all physical processes~\cite{Ivanov:2015tru}.

Finally, the matrix elements in Eqs.~(\ref{eq:VP}) and~(\ref{eq:VV}) are written down as linear combinations
of the Lorentz structures multiplied by the scalar functions--form factors which depend on the momentum transfer squared.
For the $V \to P$  transition one has
 \begin{align}
&\langle D^-_{(s)} (p_2) \left|\bar{q} O_{\mu}  c \right| 
J/ \psi (\epsilon_1 , p_1)\rangle 
\nn
&=\frac{\epsilon_1^{\nu}}{m_1 + m_2}
[ - g_{\mu\nu}pqA_0(q^2) + p_{\mu}p_{\nu}A_+(q^2)+q_{\mu}p_{\nu}
A_-(q^2) + i \varepsilon_{\mu\nu\alpha\beta}p^{\alpha}q^{\beta}V(q^2)],
\label{eq:FFpseudoscalar}
\end{align}
where $q=p_1-p_2$, $p=p_1+p_2$, $m_1\equiv m_{J/\psi}$, $m_2\equiv m_{D_{(s)}}$.

For comparison of results we relate our form factors to those defined, 
e.g., in~\cite{Khodjamirian:2006st}, which are denoted by a superscript $c$. 
The relations read
\be
\begin{array}{c@{\qquad} l}
A_+ = A^c_2, & \displaystyle A_0 = \frac{m_1 + m_2}{m_1 - m_2}A^c_1,\\ \\
V = V^c, &\displaystyle A_- = \frac{2m_2(m_1+m_2)}{q^2}(A^c_3-A^c_0).
\end{array}
\label{eq: ffrelation}
\en
We note in addition that the form factors $A^c_i(q^2)$ satisfy the constraints 
\begin{align}
A^c_0(0) = A^c_3(0)\quad \text{and}\quad
2m_2 A^c_3(q^2) = (m_1+m_2)A^c_1(q^2)-(m_1-m_2)A^c_2(q^2)
\label{eq:FFconstraints}
\end{align}
to avoid the singularity at $q^2=0$.

In the case of $V \to V$ transition we follow the authors in~\cite{Wang:2007ys} 
and define the form factors as follows:
\begin{align}
&\langle D^{*-}_{(s)} (\epsilon_2, p_2) 
\left|\bar{q} O_\mu c \right| J/ \psi (\epsilon_1 , p_1)\rangle 
\nn
&= \varepsilon_{\mu\nu\alpha\beta} \epsilon^{\alpha}_1 \epsilon^{*\beta}_2 
\left[ \left( p^{\nu} - \frac{m^2_1 - m^2_2}{q^2} q^{\nu}\right) A_1(q^2) 
+ \frac{m^2_1 - m^2_2}{q^2} q^{\nu} A_2(q^2) \right] 
\nn
& + \frac{i}{m^2_1 - m^2_2} \varepsilon_{\mu\nu\alpha\beta} p^{\alpha}_1 p^{\beta}_2 
\left[ A_3(q^2) \epsilon^{\nu}_1 \epsilon^*_2 \cdot q - A_4(q^2) \epsilon^{*\nu}_2 
\epsilon_1 \cdot q \right] 
\nn
&+ (\epsilon_1 \cdot \epsilon^*_2) \left[ -p_{\mu} V_1(q^2) + q_{\mu} V_2(q^2)
\right]
\nn
&+ \frac{(\epsilon_1 \cdot q)(\epsilon^*_2 \cdot q)}{m^2_1 - m^2_2} 
\left[ \left(p_{\mu}  - \frac{m^2_1 - m^2_2}{q^2} q_{\mu}\right) V_3(q^2) 
+ \frac{m^2_1 - m^2_2}{q^2} q_{\mu} V_4(q^2) \right] 
\nn
&- (\epsilon_1 \cdot q) {\epsilon^*_2}_{\mu} V_5(q^2) 
+ (\epsilon^*_2 \cdot q) {\epsilon_1}_{\mu} V_6(q^2).
\label{eq:FFvector}
\end{align}

The form factors in our model are represented by the threefold integrals
which are calculated by using \textsc{fortran} codes in the full kinematical momentum 
transfer region.   

\section{Form factors}
\label{sec:FF}

Before listing our numerical results we need to specify parameters of 
the CCQM that cannot be evaluated from first principles. They are the 
size parameter of hadrons $\Lambda$, the universal infrared cutoff parameter 
$\lambda$ and the constituent quark masses $m_{q_i}$. These parameters are 
determined by a least-squares fit of calculated meson leptonic decay 
constants and several fundamental electromagnetic decays to experimental data 
and/or lattice simulations within a root-mean-square deviation of $15\%$~\cite{Ivanov:2000aj}. 
This value can provide a reasonable estimate of our theoretical error since 
the calculations in our work are, in principle, not different from those 
used in the fit. For example, based on a widespread application in 
a previous paper~\cite{Ivanov:2007cw}, we suggested that a reasonable 
estimate of our theoretical error is $15\%$. 

The most recent fit results for those parameters involved in this paper 
are given in (\ref{eq:modelparam}) (all in GeV):
\be
\begin{tabular}{ c c c  c  c c c c c  }
\quad $m_{u/d}$ \quad & \quad $m_s$ \quad & \quad $m_c$ \quad
& \quad $\lambda$ \quad & \quad 
$\Lambda_{J/\psi}$ \quad & \quad $\Lambda_{D^*}$ \quad & \quad 
$\Lambda_{D^*_s}$ \quad & \quad $\Lambda_D$ \quad  
& \quad $\Lambda_{D_s}$ \quad 
\\
 \quad 0.241 \quad & \quad 0.428 \quad & \quad 1.67 \quad & \quad 0.181 \quad 
& \quad 1.74 \quad & \quad 1.53 \quad & \quad 1.56 \quad & \quad 1.60 \quad & 
\quad 1.75 \quad  \\
\end{tabular}
.
\label{eq:modelparam}
\en
Model-independent parameters and other physical constants like the Cabibbo-Kobayashi-Maskawa matrix 
elements, mass and decay width of the particles are taken 
from~\cite{Agashe:2014kda}. For clarity we note that we use the values 
$|V_{cd}|=0.225$ and $|V_{cs}|=0.986$.
\begin{table}[htbp]
\caption{Results for the leptonic decay constants $f_H$ in MeV. }
\setlength{\tabcolsep}{11pt}
\renewcommand{\arraystretch}{1}
\begin{tabular}{l l c c}
\hline\hline
{} & This work & Other & Reference\\
\hline
$f_{J/\psi}$ & 415.0 & 418$\pm$9 & LAT and QCD SR~\cite{Becirevic:2013bsa} 
\\
$f_D$ & 206.1  & 204.6$\pm$5.0 & PDG~\cite{Agashe:2014kda} 
\\
$f_{D^*}$ & 244.3 & $245(20)^{+3}_{-2}$ & LAT~\cite{Becirevic:1998ua}
\\
{} & {} & $278\pm13\pm10$ & LAT~\cite{Becirevic:2012ti} 
\\
{} & {} & $252.2\pm22.3\pm4$ & QCD SR~\cite{Lucha:2014xla}
\\
$f_{D_s}$ & 257.5 & 257.5$\pm$4.6 & PDG~\cite{Agashe:2014kda}
\\
$f_{D^*_s}$ & 272.0 & $272(16)^{+3}_{-20}$ & LAT~\cite{Becirevic:1998ua}
\\
{} & {} & 311$\pm$9 & LAT~\cite{Becirevic:2012ti}
\\
{} & {} & $305.5\pm26.8\pm5$ & QCD SR~\cite{Lucha:2014xla}
\\
$f_{D_s}/f_D$ & 1.249 & 1.258$\pm$0.038 & PDG~\cite{Agashe:2014kda}
\\
$f_{D^*_s}/f_{D^*}$ & 1.113 & 1.16$\pm$0.02$\pm$0.06 & LAT~\cite{Becirevic:2012ti}
\\
\hline\hline
\end{tabular}
\label{tab:decayconst}
\end{table} 

We present our results for the leptonic decay constants of the $J/\psi$ and 
$D^{(*)}_{(s)}$ mesons in Table~\ref{tab:decayconst}. We also list the values 
of these constants obtained from experiments or other theoretical studies 
for comparison.
One can see that our calculated values are consistent (within $10\%$) 
with results of other studies.

\begin{figure}[htbp]
\begin{tabular}{lr}
\includegraphics[width=0.50\textwidth]{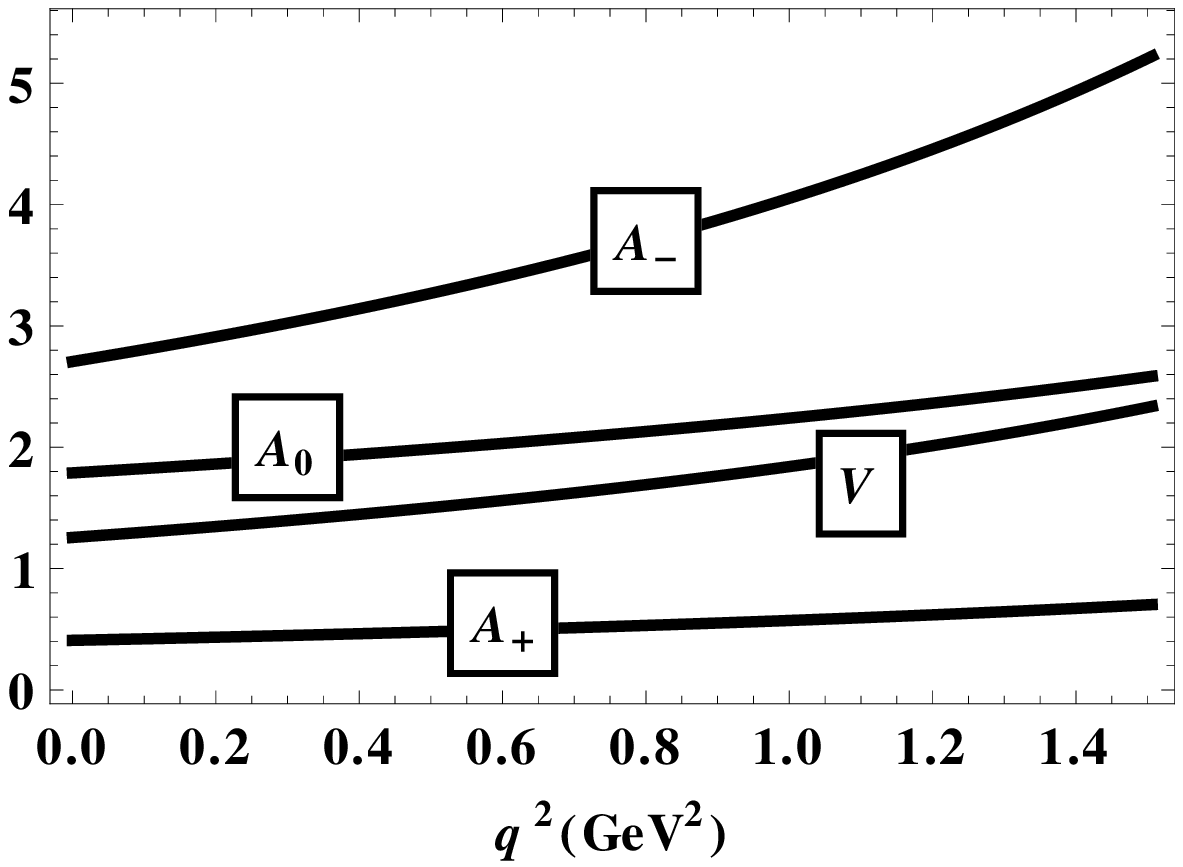} &
\includegraphics[width=0.50\textwidth]{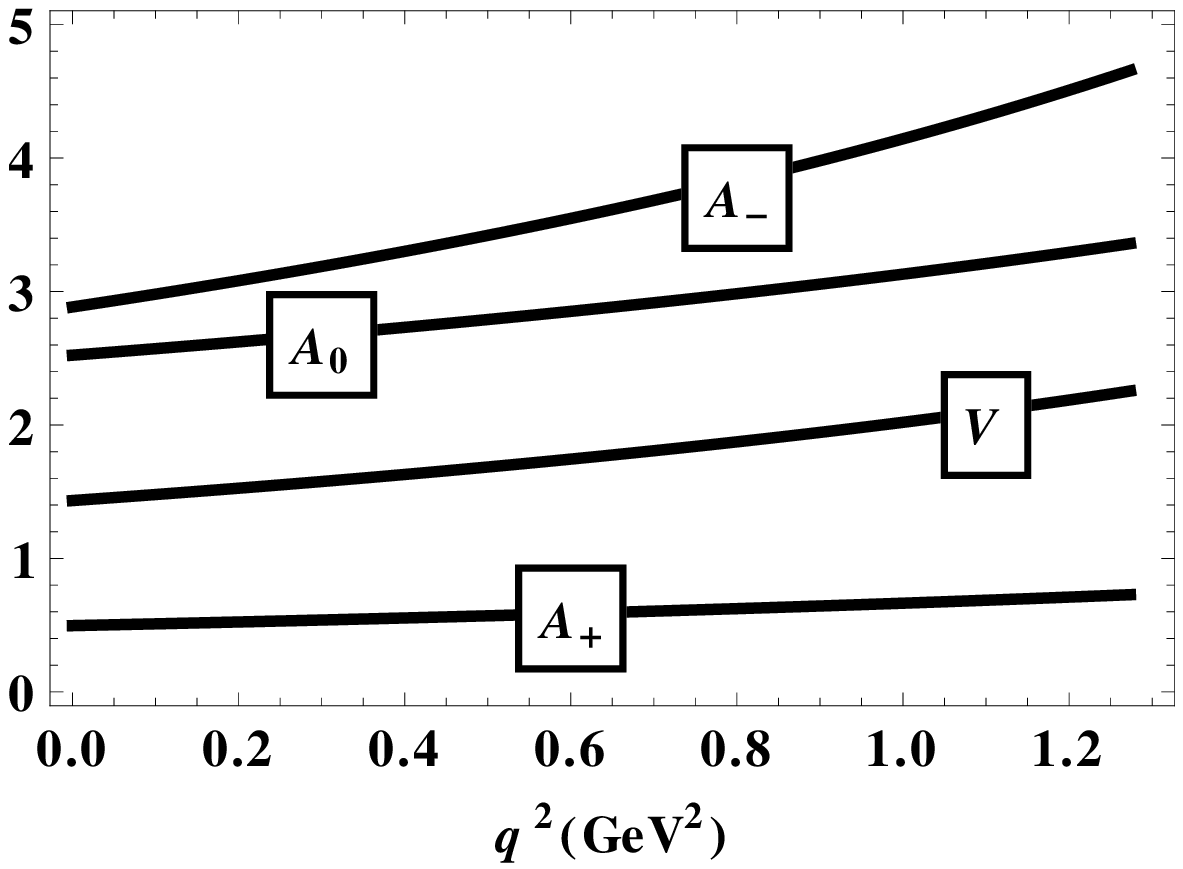}
\end{tabular}
\caption{Our results for the form factors of the $J/\psi \to D$ (left) and 
$J/\psi \to D_s$ (right) transitions.} 
\label{fig:FF-Ds}
\end{figure}

\begin{figure}[htbp]
\begin{tabular}{lr}
\includegraphics[width=0.50\textwidth]{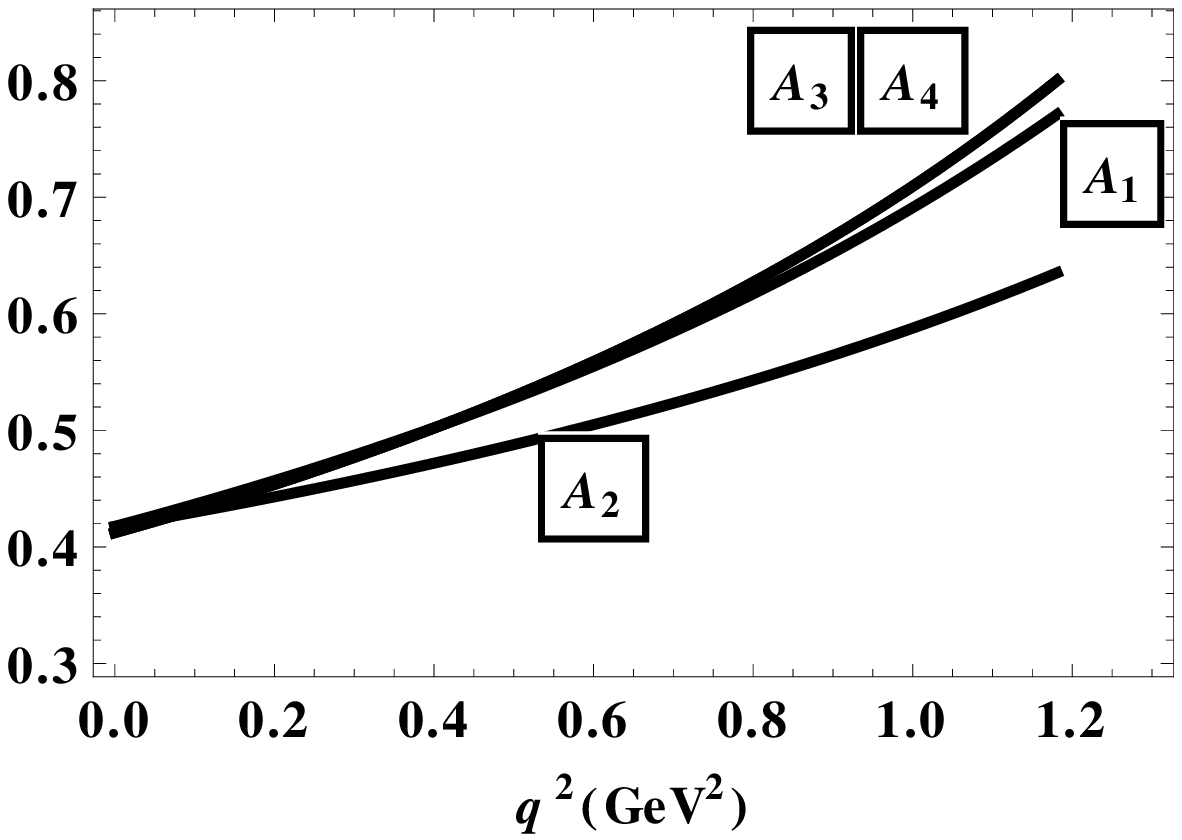} &
\includegraphics[width=0.50\textwidth]{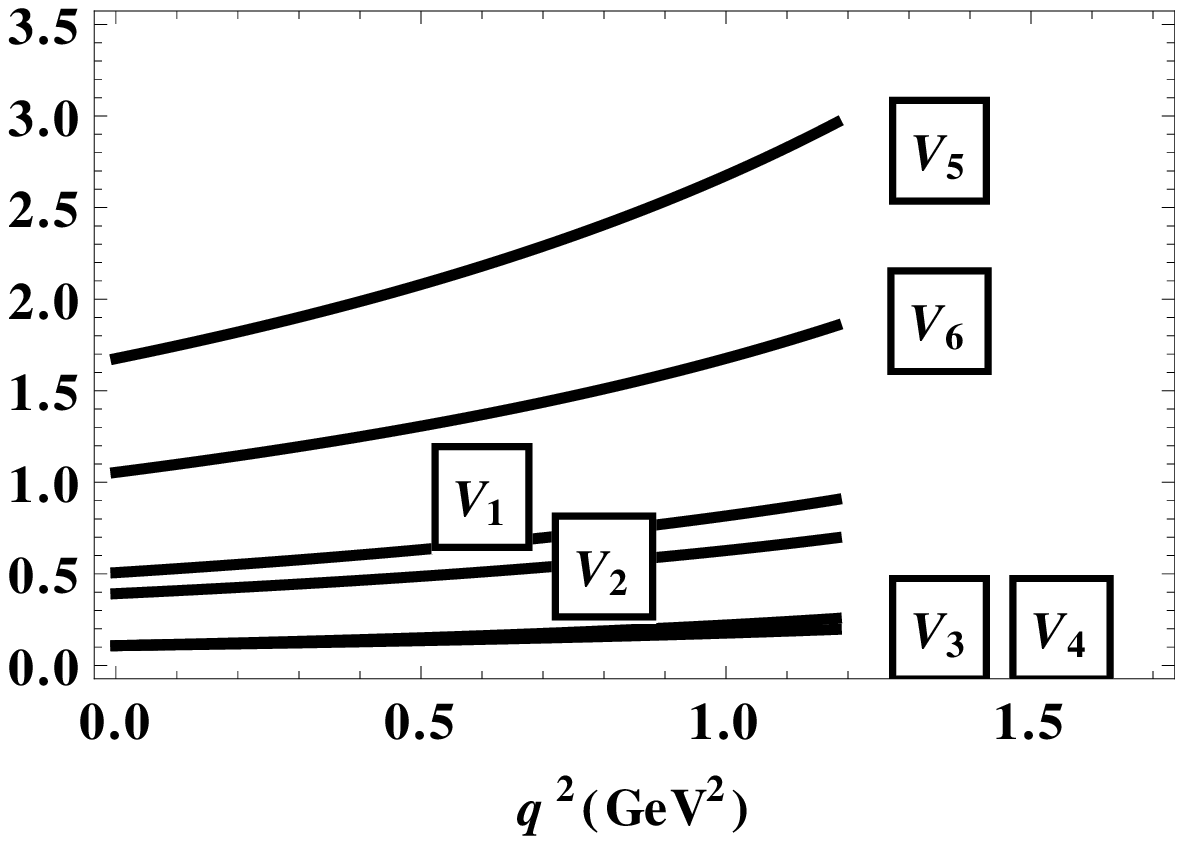}
\end{tabular}
\caption{Our results for the form factors of the $J/\psi \to D^*$ transition. 
One has to note that in the left panel $A_1(0)=A_2(0)$ and 
$A_3(q^2)\equiv A_4(q^2)$.}
\label{fig:FF-Dv}
\end{figure}

\begin{figure}[htbp]
\begin{tabular}{lr}
\includegraphics[width=0.50\textwidth]{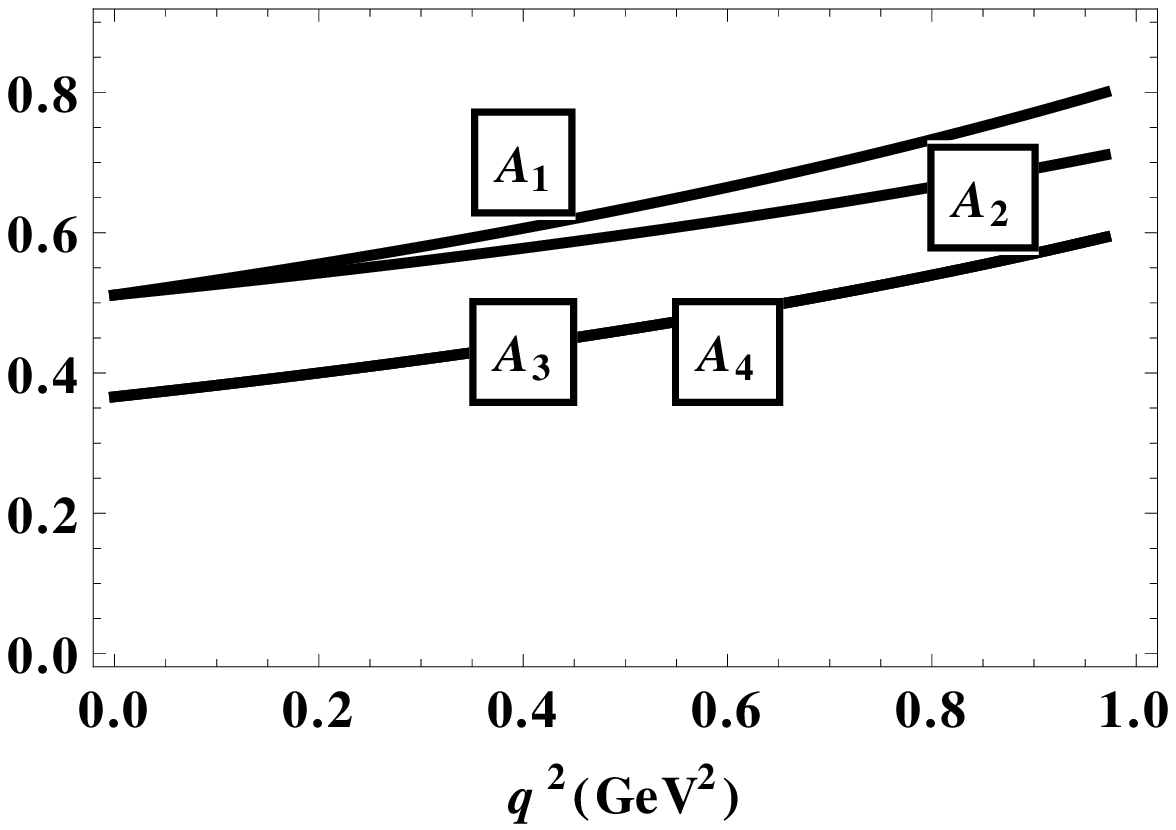} &
\includegraphics[width=0.50\textwidth]{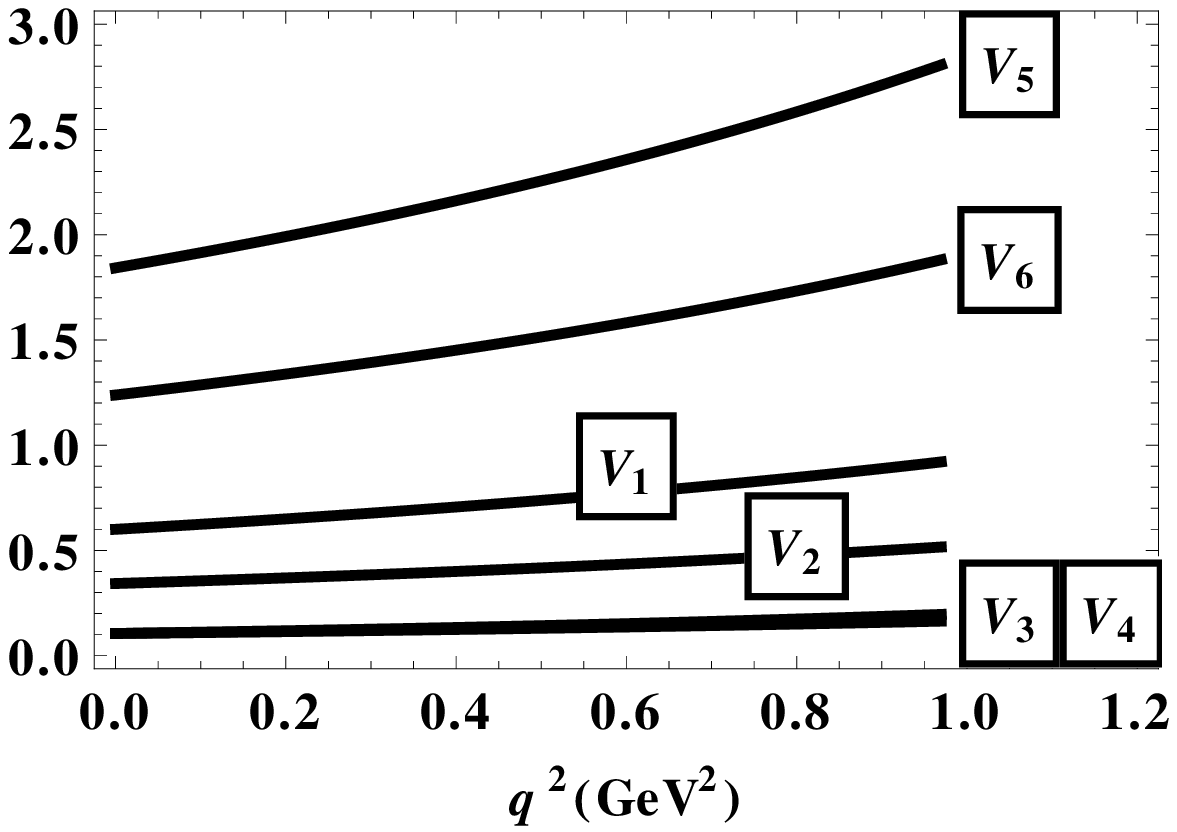}
\end{tabular}
\caption{Our results for the form factors of the $J/\psi \to D^*_s$ 
transition. One has to note that in the left panel 
$A_1(0)=A_2(0)$ and $A_3(q^2)\equiv A_4(q^2)$.}
\label{fig:FF-Dvs}
\end{figure}

In Fig.~\ref{fig:FF-Ds}-\ref{fig:FF-Dvs} we present the $q^2$ dependence 
of calculated form factors of the $J/\psi \to D^{(*)}_{(s)}$ transitions 
in the full range of momentum transfer 
$0\le q^2 \le q^2_{\rm max} = (m_{J/\psi}-m_{D^{(*)}_{(s)}})^2$. 
We found that the form factors $A_3$ and $A_4$ defined in~(\ref{eq:FFvector})
are very similar to each other.
As mentioned earlier, the CCQM allows one to evaluate form factors 
in the full kinematical range including the near-zero recoil region. 
This feature is one of those that distinguish the CCQM from other 
frameworks like QCD SR and some other approaches.
For example, the physical region of $q^2$ for  
$J/\psi \to D^- {\ell}^+ \nu_{\ell}$ is 
$0 \le q^2 \le (m_{J/\psi}-m_{D^-})^2 \simeq 1.51 \,\text{GeV}^2$. 
However, within the QCD SR approach, the authors of~\cite{Wang:2007ys} 
had to restrict their calculations in the range of 
$q^2 \in [0,\,0.47]\,\text{GeV}^2$ to avoid additional singularities 
and then use an extrapolation to obtain the form factors in large $q^2$ region.
 As a result, the extrapolation type becomes more sensitive.

\begin{figure}[htbp]
\includegraphics[width=0.60\textwidth]{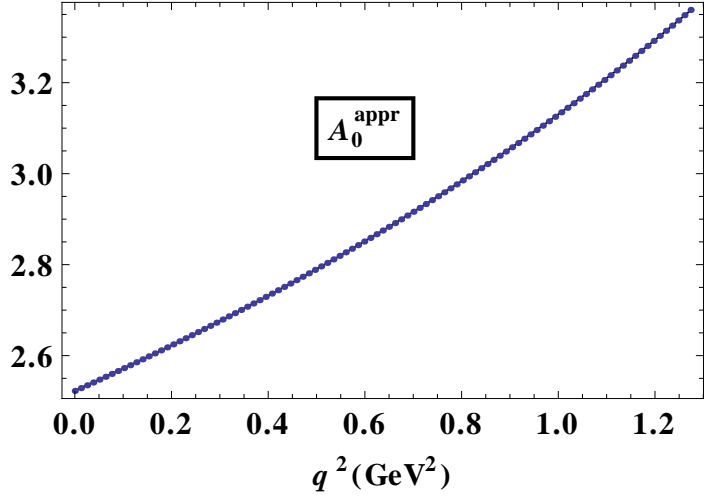} 
\caption{Comparison of $A_0(q^2)$ form factor for the $J/\psi\to D_s$ transition
calculated by \textsc{fortran} code (dotted) with parametrization given by
Eq.~(\ref{eq:DPP}) (solid).}
\label{fig:FF-comp}
\end{figure}

The results of our numerical calculation are well represented
by a double-pole parametrization
\be
F(q^2)=\frac{F(0)}{1 - a s + b s^2}, \quad s=\frac{q^2}{m_1^2},  
\label{eq:DPP}
\en 
where $m_1 = m_{J/\psi}$. The double-pole approximation is quite accurate. 
The relative error relative to the exact results is less than $1\%$ over 
the entire $q^{2}$ range, as demonstrated in Fig.~\ref{fig:FF-comp}. 


For the $J/\psi \to D_{(s)}^{(\ast)}$ transitions the parameters of the dipole
approximation are displayed in Tables~\ref{tab:FF-JD-param} and 
\ref{tab:FF-JDv-param}.
\begin{table}[ht]
\caption{Parameters of the dipole approximation for $J/\psi\to D_{(s)}$
form factors.}
\setlength{\tabcolsep}{11pt}
\renewcommand{\arraystretch}{0.9}
\begin{center}
\begin{tabular}{c|cccc|cccc}
\hline\hline
\multicolumn{1}{c|}{} &\multicolumn{4}{c|}{$J/\psi\to D$} 
                      &\multicolumn{4}{c}{$J/\psi\to D_s$} \\
\hline
 & $ A_0 $ & $  A_+  $ & $  A_-  $ & $  V  $ 
 & $ A_0 $ & $  A_+  $ & $  A_-  $ & $  V  $ 
 \\
\hline
$F(0)$ &  1.79    & 0.41 & 2.71 & 1.26 & 2.52   & 0.50 &  2.88 & 1.43
\\
$a$    &  1.87    & 2.90 & 3.41 & 3.24 & 1.81   & 2.53 &  3.10 & 2.94
\\[1ex] 
$b$    &  $-0.56$ & 1.43 & 2.21 & 1.89 &$-0.47$ & 0.98 &  1.76 & 1.48
\\ 
\hline\hline
\end{tabular}
\label{tab:FF-JD-param}
\end{center}
\end{table}
\begin{table}[ht]
\caption{Parameters of the dipole approximation for $J/\psi\to D^\ast_{(s)}$
form factors.}
\setlength{\tabcolsep}{10pt}
\renewcommand{\arraystretch}{0.9}
\begin{center}
\begin{tabular}{c|cccccccccc}
\hline\hline
\multicolumn{1}{c|}{} &\multicolumn{10}{c}{$J/\psi\to D^\ast$} \\ 
\hline
 & $ A_1 $ & $  A_2  $ & $  A_3  $ & $  A_4 \qquad $ 
 & $ V_1 $ & $  V_2  $ & $  V_3  $ & $  V_4  $ & $  V_5  $ & $  V_6  $  
 \\
\hline
$F(0)$ & 0.42 & 0.42    & 0.41 & 0.41 & 0.51 & 0.39 & 0.11 &  0.11 & 1.68 & 1.05
\\
$a$    & 4.20 & 2.75    & 4.46 & 4.46 & 3.98 & 3.85 & 4.03 &  6.00 & 3.88 & 3.85
\\[1ex] 
$b$    & 3.87 & $-0.30$ & 4.27 & 4.27 & 3.25 & 2.44 & 2.95 & 10.56 & 2.83 & 2.80
\\ 
\hline
\multicolumn{1}{c|}{} &\multicolumn{10}{c}{$J/\psi\to D_s^\ast$} \\ 
\hline
$F(0)$ & 0.51 & 0.51   & 0.37 & 0.37 & 0.60 & 0.34 & 0.10 & 0.10 & 1.84  & 1.23 
\\
$a$    & 3.89 &  2.76  & 4.15 & 4.15 & 3.72 & 3.52 & 3.80 & 5.46 & 3.64  & 3.62
\\[1ex] 
$b$    & 3.15 &$-0.18$ & 3.57 & 3.57 & 2.72 & 1.94 & 2.53 & 8.82 & 2.39  & 2.37
\\ 
\hline\hline
\end{tabular}
\label{tab:FF-JDv-param}
\end{center}
\end{table}


In Tables \ref{tab:FF-JD-comp} and \ref{tab:FF-JDv-comp}
we compare the values of our form factors at $q^2=0$ (maximum recoil)
with those obtained within QCD SR~\cite{Wang:2007ys} and 
LFQM~\cite{Shen:2008zzb}. Our results are more consistent with those 
in~\cite{Wang:2007ys}. For example, our predictions for the form factors at 
$q^2=0$ differ from results of~\cite{Wang:2007ys} within $40\%$ while 
the discrepancy can come to a factor of $4$ comparing with the results 
of~\cite{Shen:2008zzb}. 
\begin{table}[ht]
\caption{Comparison of $J/\psi \to D_{(s)}$ form factors at maximum
recoil with those obtained in QCD SR and LFQM.}
\setlength{\tabcolsep}{12pt}
\renewcommand{\arraystretch}{0.9}
\begin{center}
\begin{tabular}{c|cccc|cccc}
\hline\hline
\multicolumn{1}{c|}{} &\multicolumn{4}{c|}{$J/\psi\to D:\quad q^2=0$} 
                      &\multicolumn{4}{c}{$J/\psi\to D_s:\quad q^2=0$} \\
\hline
 & $ A_0 $ & $  A_+  $ & $  A_-  $ & $  V $ 
 & $ A_0 $ & $  A_+  $ & $  A_-  $ & 
$  V  $ 
 \\
\hline
QCD SR~\cite{Wang:2007ys} 
               & 1.09 & 0.34 & $\dots$  & 0.81 & 1.71  & 0.35 & $\dots$  & 1.07
\\[1ex] 
LFQM~\cite{Shen:2008zzb}   
               & 2.75 & 0.18 & $\dots$  & 1.6  & 3.05  & 0.13 & $\dots$  & 1.8
\\
Our results            & 1.79 & 0.41 & 2.71 & 1.26 & 2.52  & 0.50 & 2.88 & 1.43
\\
\hline\hline
\end{tabular}
\label{tab:FF-JD-comp}
\end{center}
\end{table}
\begin{table}[htbp]
\caption{Comparison of $J/\psi \to D_{(s)}^\ast$ form factors at 
maximum recoil with those obtained in QCD SR~\cite{Wang:2007ys}.}
\setlength{\tabcolsep}{10pt}
\renewcommand{\arraystretch}{0.9}
\begin{center}
\begin{tabular}{c|cccccccccc}
\hline\hline
\multicolumn{1}{c|}{} &
\multicolumn{10}{c}{$J/\psi\to D^\ast:\quad q^2=0$} \\ 
\hline
 & $ A_1 $ & $  A_2  $ & $  A_3  $ & 
$  A_4  $ & 
$ V_1 $ & $  V_2  $ & $  V_3  $ & 
$  V_4  $ & $  V_5  $ & $  V_6  $  
 \\
\hline
\cite{Wang:2007ys}    
    & 0.40 & 0.44 & 0.86 & 0.91 & 0.41 & 0.63 & 0.22 &  0.26 & 1.37 & 0.87
\\ 
Our results & 0.42 & 0.42 & 0.41 & 0.41 & 0.51 & 0.39 & 0.11 &  0.11 & 1.68 & 1.05
\\
\hline
\multicolumn{1}{c|}{} &
\multicolumn{10}{c}{$J/\psi\to D_s^\ast:\quad q^2=0$} \\ 
\hline
\cite{Wang:2007ys}    
    & 0.53 & 0.53 & 0.91 & 0.91 & 0.54 & 0.69 & 0.24 & 0.26 & 1.69 & 1.14
\\ 
Our results & 0.51 & 0.51 & 0.37 & 0.37 & 0.60 & 0.34 & 0.11 & 0.11 & 1.84  & 1.24  
\\
\hline\hline
\end{tabular}
\label{tab:FF-JDv-comp}
\end{center}
\end{table}

\section{Numerical results}
\label{sec:branching}
The invariant matrix element for the decay
$J/\psi \to D_{(s)}^{(\ast)\,-} \ell^+ \nu_{\ell}$
is written down as
\be
\mathcal{M}=
\frac{G_F}{\sqrt{2}}V_{cq}
\left\langle D^-\left|\bar{q} O_{\mu}  c \right| J/ \psi \right\rangle
\left[\bar{\nu_{\ell}}O^{\mu}\ell\right]. 
\en 
The unpolarized lepton tensor for the process
       $W^-_{\rm off-shell}\to \ell^-\bar \nu_\ell$ 
$\left( W^+_{\rm off-shell}\to \ell^+ \nu_\ell \right)$  
is given  by \cite{Gutsche:2015mxa}
\bea
L^{\mu\nu} &=& \left\{\begin{array}{lr}
\Tr\Big[ (\slp_\ell + m_\ell) O^\mu \slp_{\nu_\ell} O^\nu\Big]
& \qquad\text{for}\qquad  W^-_{\rm off-shell}\to \ell^-\bar\nu_\ell 
\\[1.2ex]
\Tr\Big[ (\slp_\ell - m_\ell) O^\nu \slp_{\nu_\ell} O^\mu\Big]
& \qquad\text{for}\qquad  W^+_{\rm off-shell}\to \ell^+ \nu_\ell 
                   \end{array}\right.
\nn[1.2ex]
&=&
8 \left( 
  p_\ell^\mu p_{\nu_\ell}^\nu  + p_\ell^\nu p_{\nu_\ell}^\mu 
- p_{\ell}\cdot p_{\nu_\ell}g^{\mu\nu}
  \pm  i \varepsilon^{\mu \nu \alpha \beta} p_{\ell\alpha} p_{\nu_\ell\beta}
\right),
\label{eq:lept_tensor}
\ena
where the upper/lower sign 
refers to the two $(\ell^-\bar\nu_\ell)/(\ell^+\nu_\ell)$ configurations.
The sign change can be seen to result from the parity violating part of 
the lepton tensors. In our case we have to use the lower sign
in Eq.~(\ref{eq:lept_tensor}).
Summing up the vector polarizations, one finds the decay rate
\be
\Gamma\left (J/\psi \to D_{(s)}^{(\ast)-} \ell^+ \nu_{\ell} \right)
= \frac{G_F^2}{(2\pi)^3}\frac{|V_{cq}|^2}{64m_1^3}
\int\limits_{m^2_{\ell}}^{(m_1-m_2)^2}\!\!\!\! dq^2
\int\limits_{s_1^-}^{s_1^+}\!\! ds_1
\frac13 H_{\mu\nu} L^{\mu\nu}.
\label{eq:rate}
\en
Here $m_1=m_{J/\psi}$, $m_2=m_D$, and $s_1 =(p_D+p_{\ell})^2$. 
The upper and lower bounds of $s_1$ are given by
\be
s_1^{\pm}=
m_2^2+m_{\ell}^2-\frac{1}{2q^2}
\left[(q^2-m_1^2+m_2^2)(q^2+m_{\ell}^2)
      \mp\lambda^{1/2}(q^2,m_1^2,m_2^2)\lambda^{1/2}(q^2,m_{\ell}^2,0)\right],
\en
where $\lambda(x,y,z) \equiv x^2+y^2+z^2-2(xy+yz+zx)$ is 
the K{\"a}ll{\'e}n function. 

The hadron tensor reads
\be
H_{\mu\nu} = \left\{\begin{array}{lr}
T^{\rm VP}_{\mu\alpha}\left(-g^{\alpha\alpha'}+\frac{p_1^\alpha p_1^{\alpha'}}{m_1^2}\right)
T^{\rm VP\dagger}_{\nu\alpha'} 
& \quad\text{for}\quad V\to P~\text{transition}
\\[1.2ex]
T^{\rm VV}_{\mu\alpha\beta}
\left(-g^{\alpha\alpha'}+\frac{p_1^\alpha p_1^{\alpha'}}{m_1^2}\right)
\left(-g^{\beta\beta'}+\frac{p_2^\beta p_2^{\beta'}}{m_2^2}\right)
T^{\rm VV\dagger}_{\nu\alpha'\beta'} 
& \quad\text{for}\quad V\to V~\text{transition}.
\end{array}\right.
\en
\begin{table}[htbp]
\caption{Semileptonic decay branching fractions of $J/\psi$ meson.}
\setlength{\tabcolsep}{12pt}
\renewcommand{\arraystretch}{0.9}
\begin{tabular}{c c c c c c}
\hline\hline
Mode & Unit &This work & QCD SR~\cite{Wang:2007ys} & LFQM~\cite{Shen:2008zzb}\\
\hline
$J/\psi \to D^- e^+ \nu_e$ & $10^{-12}$ & 17.1  & $7.3^{+4.3}_{-2.2}$ 
& $51 - 57$\\
$J/\psi \to D^- \mu^+ \nu_{\mu}$ & $10^{-12}$ & 16.6  & $7.1^{+4.2}_{-2.2}$ 
& $47 - 55$ \\
$J/\psi \to D^-_s e^+ \nu_e$ & $10^{-10}$ & 3.3 & $1.8^{+0.7}_{-0.5}$ 
& $5.3 - 5.8$ \\
$J/\psi \to D^-_s \mu^+ \nu_{\mu}$ & $10^{-10}$ & 3.2 & $1.7^{+0.7}_{-0.5}$ 
& $5.5 - 5.7$ \\
$J/\psi \to D^{*-} e^+ \nu_e$ & $10^{-11}$ & 3.0 & $3.7^{+1.6}_{-1.1}$ 
& $\dots$ \\
$J/\psi \to D^{*-} \mu^+ \nu_{\mu}$ & $10^{-11}$ & 2.9 & $3.6^{+1.6}_{-1.1}$ 
& $\dots$ \\
$J/\psi \to D^{*-}_s e^+ \nu_e$ & $10^{-10}$ & 5.0 & $5.6^{+1.6}_{-1.6}$ 
& $\dots$\\
$J/\psi \to D^{*-}_s \mu^+ \nu_{\mu}$ & $10^{-10}$ & 4.8 & $5.4^{+1.6}_{-1.5}$ 
& $\dots$\\
\hline\hline
\end{tabular}
\label{tab:branching}
\end{table}

We present our results for the branching fractions in 
Table~\ref{tab:branching} together with results of other theoretical studies 
based on QCD SR and LFQM for comparison. It is worth mentioning that all 
values for $\mathcal{B}(J/\psi \to D^*_{(s)} \ell \nu)$ are fully consistent 
with those in~\cite{Wang:2007ys}. 
Regarding $\mathcal{B}(J/\psi \to D_{(s)} \ell \nu)$, our results are larger 
than those in~\cite{Wang:2007ys} by a factor of $2-3$. We think this 
discrepancy is mainly due to the values of the meson leptonic decay constants 
$f_D=166\,\text{MeV}$ and $f_{D_s}=189\,\text{MeV}$ used in~\cite{Wang:2007ys}, 
which are much smaller than $f_D=206.1\,\text{MeV}$ and 
$f_{D_s}=257.5\,\text{MeV}$ used in our present paper. In contrast, 
the constants $f_{D^*}=240\,\text{MeV}$ and $f_{D^*_s}=262\,\text{MeV}$ used 
in~\cite{Wang:2007ys} are very close to our values of $f_{D^*}=244.3\,\text{MeV}$ 
and $f_{D^*_s}=272.0\,\text{MeV}$, resulting in a full agreement in 
$\mathcal{B}(J/\psi \to D^*_{(s)} \ell \nu)$ between the two studies. 
Comparing with another study, our results for 
$\mathcal{B}(J/\psi \to D_{(s)} \ell \nu)$ are smaller than 
those in~\cite{Shen:2008zzb} by a factor of $2-3$.

It is interesting to consider the ratio 
$R\equiv \mathcal{B}(J/\psi \to D^*_s \ell \nu)
/\mathcal{B}(J/\psi \to D_s \ell \nu)$, where a large part of theoretical and
experimental uncertainties cancels. We list in~(\ref{eq:R}) all available 
predictions for $R$ up till now:
\be
R\equiv \frac{\mathcal{B}(J/\psi \to D^*_s \ell \nu)}
             {\mathcal{B}(J/\psi \to D_s \ell \nu)}
=\left\{\begin{array}{lc}
1.5 \qquad & \qquad \text{{M.A.~Sanchis-Lonzano}~\cite{SanchisLozano:1993ki}} \\
3.1 \qquad & \qquad \text{{Y.M.~Wang}~\cite{Wang:2007ys}}  \\
1.5  \qquad & \qquad \text{This work}
\end{array}\right.
.
\label{eq:R}
\en
Wang's result for $R$ is about two times greater than our prediction because 
their branching fraction $\mathcal{B}(J/\psi \to D_s \ell \nu)$ is about two 
times smaller than ours (mainly due to the leptonic decay constants). 
Therefore, we propose that the value $R\simeq 1.5$ is a reliable prediction. 

Moreover, we also consider the ratios
\be
R_1 \equiv \frac{\mathcal{B}(J/\psi \to D_s \ell \nu)}
                {\mathcal{B}(J/\psi \to D \ell \nu)}
 \qquad \text{and} \qquad 
R_2 \equiv \frac{\mathcal{B}(J/\psi \to D^*_s \ell \nu)}
                {\mathcal{B}(J/\psi \to D^* \ell \nu)},
\en
which should be equal to $\displaystyle \frac{|V_{cs}|^2}{|V_{cd}|^2}\simeq 18.4$
under the $SU(3)$ flavor symmetry limit. These ratios are $R_1\simeq 24.7$ and 
$R_2\simeq 15.1$ in~\cite{Wang:2007ys}. In this work we have the following 
values, $R_1\simeq 19.3$ and $R_2\simeq 16.6$, which suggest a relative small 
$SU(3)$ symmetry breaking effect.

\section{Summary and conclusions}
\label{sec: summary}

Let us summarize the main results of our paper. 
We have calculated the hadronic form factors relevant to the 
semileptonic decay $J/\psi \to D_{(s)}^{(*)-} {\ell}^+ \nu_{\ell}$
in the framework of the confined covariant quark model.
By using the calculated form factors and Standard Model parameters
we have evaluated the decay rates and branching fractions.
We have compared our results with those obtained in other approaches. 

\begin{acknowledgments}

M.A.I.\ acknowledges Mainz Institute for Theoretical 
Physics (MITP) and the Heisenberg-Landau Grant for the support.

\end{acknowledgments}

\end{document}